\pdfoutput=1
\documentclass[10pt,twocolumn,english,sort,twocolumn]{IEEEtran}
\usepackage{helvet}

\usepackage[T1]{fontenc}
\usepackage[latin9]{inputenc}
\usepackage[letterpaper]{geometry}
\usepackage{color}
\usepackage{babel}
\usepackage{array}
\usepackage{units}
\usepackage{graphicx}
\usepackage{esint}
\usepackage[unicode=true,pdfusetitle,
 bookmarks=true,bookmarksnumbered=false,bookmarksopen=false,
 breaklinks=false,pdfborder={0 0 0},backref=false,colorlinks=false]
 {hyperref}

\makeatletter

\providecommand{\tabularnewline}{\\}

\renewcommand{\fnum@figure}{Fig.~\thefigure}

\makeatother

\begin{document}

\title{Non-Invertibility of spectral x-ray photon counting data with pileup}

\author{Robert E. Alvarez\thanks{\protect\href{http://www.aprendtech.com}{Aprendtech.com}, ralvarez@aprendtech.com}}
\maketitle
\begin{abstract}
In the Alvarez-Macovski method {[}R.E. Alvarez and A. Macovski, Phys.
Med. Biol., 1976, 733-744{]}, the attenuation coefficient is approximated
as a linear combination of functions of energy multiplied by coefficients
that depend on the material composition at points within the object.
The method then computes the line integrals of the basis set coefficient
from measurements with different x-ray spectra. This paper shows that
the transformation from photon counting detector data with pileup
to the line integrals can become ill-conditioned under some circumstances
leading to highly increased noise.

Methods: An idealized model that includes pileup and quantum noise
is used. The noise variance of the line integral estimates is computed
using the Cramèr-Rao lower bound (CRLB). The CRLB is computed as a
function of object thickness for photon counting detector data with
three and four bin pulse height analysis (PHA) and low and high pileup. 

Results: With four bin PHA data the transformation is well conditioned
with either high or low pileup. With three bin PHA and high pileup,
the transformation becomes ill-conditioned for specific values of
object attenuation. At these values the CRLB variance increases by
approximately $10^{5}$ compared with the four bin PHA or low pileup
results. The condition number of the forward transformation matrix
also shows a spike at those attenuation values.

Conclusion: Designers of systems using counting detectors should study
the stability of the line integral estimator output with their data.

\vspace{0.2cm}

\hspace{-0.1cm}Key Words: photon counting, pileup, spectral x-ray,
dual energy, spectral x-ray, energy selective, Cramèr-Rao lower bound
\end{abstract}

\section{INTRODUCTION}

This paper studies the stability of the Alvarez-Macovski method\cite{Alvarez1976}
with x-ray photon counting data with pileup. The method uses an expansion
of the x-ray attenuation coefficient as a linear combination of functions
of energy multiplied by constants that depend only on the material
composition at points in the object. Transmission measurements with
multiple x-ray spectra are then used to estimate the line integrals
of the basis set coefficients. With a photon counting detector, the
transmitted photons are analyzed with pulse height analysis\cite{Knoll2000}(PHA)
so that each bin provides a separate measurement spectrum. 

In order to focus on fundamental effects, the noise in the line integral
estimates is computed using an idealized model that only includes
pileup and quantum noise. In the computation, the transmitted spectra
are first computed using realistic models of the x-ray tube source
spectrum\cite{Boone1997} and tabulated attenuation coefficients\cite{Hubbell1999}.
The recorded photon counts with pileup are then computed using a non-paralyzable
model of the recorded photon counts with pileup\cite{Knoll2000}.
In the model, the recorded energies are assumed to be the sum of the
energies of the photons incident during the dead time period. The
recorded energy data are measured using pulse height analysis with
ideal rectangular energy bins. The PHA bin counts are the input data
to the A-vector estimator.

The probability distribution of the PHA bin counts with pileup is
modeled as multivariate normal with parameters that depend on the
spectrum of the photons incident on the detector and the pileup parameter\cite{AlvarezSNRwithPileup2014}.
A universal limit on the A-vector estimate noise variance is computed
using the Cramèr-Rao lower bound (CRLB), which is the minimum covariance
for any unbiased estimator\cite{KayV1Chapter3}. The CRLB is computed
from the multivariate normal parameters\cite{KayV1Chapter3} as a
function of object thickness for photon counting detector data with
three or four bin PHA with low and high pileup. 

The results show that with high pileup three bin PHA data the noise
variance exhibits a sharp peak at specific object thicknesses. The
peak does not appear with four bin PHA data or three bin data with
low pileup. The peak variance is approximately $10^{5}$ times the
values for the same thicknesses in the cases without the instability.

\section{METHODS}

\subsection{The Alvarez-Macovski method\label{sub:Alv-Mac-method}}

For biological materials and an externally administered high atomic
number contrast agent, we can approximate the x-ray attenuation coefficient
$\mu(\mathbf{r},E)$ accurately with a three function basis set\cite{alvarez2013dimensionality}
\begin{equation}
\mu(\mathbf{r},E)=a_{1}(\mathbf{r})f_{1}(E)+a_{2}(\mathbf{r})f_{2}(E)+a_{3}(\mathbf{r})f_{3}(E).\label{eq:3-func-decomp}
\end{equation}

\noindent{}In this equation, $a_{i}(\mathbf{r})$ are the basis set
coefficients and $f_{i}(E)$ are the basis functions, $i=1\ldots3$.
As implied by the notation, the coefficients $a_{i}(\mathbf{r})$
are functions only of the position $\mathbf{r}$ within the object
and the functions $f_{i}(E)$ depend only on the x-ray energy $E$.
If there is more than one high atomic number material present, we
can extend the basis set to higher dimensions. 

Neglecting scatter, the expected value of the number of transmitted
photons $\lambda_{k}$ with an effective measurement spectrum $S_{k}(E)$
is 
\begin{equation}
\lambda_{k}=\int S_{k}(E)e^{-\int\mu\left(\mathbf{r},E\right)d\mathbf{r}}dE\label{eq:Ik-integral}
\end{equation}

\noindent{}where the line integral in the exponent is on a line from
the x-ray source to the detector. For idealized PHA data, the effective
spectrum for each energy bin is 
\begin{equation}
S_{k}\left(E\right)=S_{incident}\left(E\right)\Pi_{k}\left(E\right)\label{eq:PHA-bin-spect}
\end{equation}

\noindent{}where $S_{incident}\left(E\right)$ is the x-ray spectrum
incident on the detector and $\Pi_{k}\left(E\right)$ is a rectangle
function equal to one inside the energy bin and zero outside. 

Using the decomposition, Eq. \ref{eq:3-func-decomp}, the line integral
in Eq. \ref{eq:Ik-integral} is 
\begin{equation}
\int\mu\left(\mathbf{r},E\right)d\mathbf{r}=A_{1}f_{1}(E)+A_{2}f_{2}(E)+A_{3}f_{3}(E).\label{eq:L(E)-A1-f1-A2-f2}
\end{equation}

\noindent{}where $A_{i}=\int a_{i}\left(\mathbf{r}\right)d\mathbf{r},\ i=1\ldots3$
are the line integrals of the basis set coefficients. The $A_{i}$
are summarized as the components of the A-vector, $\mathbf{A}$, and
the measurements by a vector, $\mathbf{N}$, whose components are
the photon counts with each effective spectrum. Since the body transmission
is exponential in $\mathbf{A}$, we can approximately linearize the
measurements by taking logarithms. The results is the log measurement
vector $\mathbf{L}=-\log(\mathbf{N/\mathbf{N_{0}}})$, where $\mathbf{N_{0}}$
is the expected value of the measurements with no object in the beam
and the division means that corresponding members of the vectors are
divided. 

Equations \ref{eq:Ik-integral} define a relationship between $\mathbf{A}$
and the expected value of the measurement vector, $\mathbf{L(A)}$.
For x-ray measurements with noise, we can use a statistical estimator
to invert the relationship and to compute the best estimate of $\hat{\mathbf{A}}$
from the measurements $\mathbf{L}$ taking into account the probability
distribution of the noise.

\subsection{The CRLB for A-vector noise\label{sub:CRLB-with-pileup}}

In general, the estimator noise depends on its implementation but
we can derive fundamental limits on x-ray system performance by using
the Cramèr-Rao lower bound (CRLB)\cite{Alvarez1976,AlvarezSNRwithPileup2014}.
The CRLB is the minimum covariance for any unbiased estimator and
is a fundamental limit from statistical estimator theory\cite{KayV1Chapter3}.
It is the inverse of the Fisher information matrix $\mathbf{F}$ whose
elements are 
\begin{equation}
F_{ij}=-\left\langle \frac{\partial^{2}\mathcal{L}}{\partial A_{i}\partial A_{j}}\right\rangle \label{eq:F-element}
\end{equation}

\noindent{}where $\mathcal{L}$ is the logarithm of the likelihood
and the symbol $\left\langle \ \right\rangle $ denotes the expected
value. 

A large number of detected photons are required for material selective
x-ray imaging and therefore we can model the measurements as having
a multivariate normal distribution\cite{AlvarezSNRwithPileup2014}.
Kay\cite{KayV1Sec4_5} shows that the Fisher matrix for multivariate
normal measurements with expected value $\mathbf{\left\langle L(A)\right\rangle }$
and covariance $\mathbf{C_{L}}$ has elements
\begin{equation}
\begin{array}{ccc}
F_{ij} & = & \left[\frac{\partial\mathbf{\left\langle L(A)\right\rangle }}{\partial A_{i}}\right]^{T}\mathbf{C_{L}^{-1}}\left[\frac{\partial\mathbf{\left\langle L(A)\right\rangle }}{\partial A_{j}}\right]+\\
 &  & \frac{1}{2}tr\left[\mathbf{C_{L}^{-1}}\frac{\partial\mathbf{C_{L}}}{\partial A_{i}}\mathbf{C_{L}^{-1}}\frac{\partial\mathbf{C_{L}}}{\partial A_{j}}\right]
\end{array}\label{eq:CRLB-gen}
\end{equation}

\noindent{}where the $tr\left[\right]$ is the trace of a matrix. 

For x-ray data, the first term in the Fisher matrix is much larger
than the second \cite{AlvarezSNRwithPileup2014}. Ignoring the second
term, the Fisher matrix is 
\begin{equation}
\mathbf{F_{const.cov.}=M^{T}C_{L}^{-1}M}\label{eq:F-const-cov}
\end{equation}

\noindent{}where $\mathbf{M}$ is the gradient matrix
\begin{equation}
\mathbf{M=}\frac{\partial\mathbf{\left\langle L(A)\right\rangle }}{\partial\mathbf{A}}\label{eq:M-dL-dA}
\end{equation}

\noindent{}with elements $M_{ij}=\nicefrac{\partial L_{i}}{\partial A_{j}}$.

Notice that $\mathbf{M}$ is also the system matrix of a linearized
first order model of the measurements about an operating point $\mathbf{\bar{A}}$
\begin{equation}
\mathbf{L}(\mathbf{A})=\mathbf{L(\overline{A})}+\mathbf{\frac{\partial L}{\partial A}(A-\overline{A})+\cdots}\label{eq:First-order-model}
\end{equation}

\noindent{}so that 
\[
\mathbf{\delta L}_{with\ noise}=\mathbf{M\delta A+w}
\]

\noindent{}\textrm{where $\mathbf{\delta L}=\mathbf{L}(\mathbf{\overline{A}+\delta A})-\mathbf{L(\overline{A})}$
, $\delta\mathbf{A=A-\overline{A}}$}, and $\mathbf{w}$ is multivariate
normally distributed measurement noise. If $\mathbf{M}$ is not invertible
then we cannot estimate the $\mathbf{A}$ vector from the spectral
measurements. In general, $\mathbf{M}$ is invertible in a mathematical
sense but, as will be shown in later sections, with pileup it can
become ill-conditioned leading to large increases in the noise variance.

\subsection{Physical model of photon counting detector data with pileup\label{sub:Models-for-pulse-pileup}}

\textcolor{black}{The idealized model for photon counting data with
pileup is described in detail in a previous paper\cite{AlvarezSNRwithPileup2014}
and is summarized here. The detector output with pileup is modeled
with the dead time parameter, $\tau$\cite{Knoll2000}, which is the
minimum time between two photons that are recorded as separate events.
Two models of pileup are commonly used. In both of the models, the
detector is assumed to start in a ``live'' state. With the arrival
of a photon, the detector enters a separate state where it does not
count additional photons. In the first model, called non-paralyzable,
the time in the separate state is assumed to be fixed and independent
of the arrival of any other photons during the dead time. In the second
model, called paralyzable, the arrival of photons extends the time
in the non-counting state. Both models give similar recorded counts
at low rates but give different results at high rates where the probability
of multiple interactions during the dead time becomes significant.
The non-paralyzable model will be used here. Measurements by Taguchi
et al.\cite{taguchi_modeling_2011} indicate it is more accurate at
higher count rates with their detectors. It also leads to simpler
analytical results\cite{yu_fessler_PMB_2000}.}

\textcolor{black}{A model is also needed for the recorded energies
with pileup. One approach is to assume that the recorded energy is
proportional to the integral of the sensor charge pulses during the
dead time\cite{taguchi_modeling_2011}. An idealization of this model
is to assume that the recorded energy is the sum of the energies of
the photons that arrive during the dead time regardless of how close
the arrival of a photon to the end of the period\cite{wang_pulse_2011}.
The idealized model assumes that the photon energy is converted completely
into charge carriers so there are no losses due to Compton or Rayleigh
scattering and all K-fluorescence radiation is re-absorbed within
the sensor. All of the carriers are assumed to be collected so there
is no charge trapping or charge sharing with nearby detectors. }

\subsection{Probability distribution of PHA data with pileup\label{sub:PHA-stats}}

The probability distribution of the PHA bin counts is modeled as multivariate
normal with the expected value and covariance in Table \ref{tab:PHA-formulas-1}.
This model can be shown to be accurate for the number of x-ray photons
required for material selective imaging\cite{AlvarezSNRwithPileup2014}.

The parameters in the Table are defined as follows: the expected value
of the total number of photons incident on the detector during the
measurement time is 
\begin{equation}
\lambda=\int S_{incident}(E)dE\label{eq:Expected-total-incident}
\end{equation}
\noindent{}where $S_{incident}(E)$ is the spectrum of the x-ray
photons incident on the detector. This spectrum can be computed for
an object with A-vector $\mathbf{A}$ using Equations \ref{eq:Ik-integral}
and \ref{eq:L(E)-A1-f1-A2-f2}. The average rate of photons incident
on the detector is $\rho$ and the pileup parameter, $\eta=\rho\tau$,
is the expected number of photons arriving during a dead time period,
$\tau$. The total number of photons recorded by the detector in all
PHA bins with pileup is $N_{rec}$ and the number recorded in PHA
bin $k$ is $N_{rec,k}$. Notice that with non-zero $\eta$, the non-Poisson
factor $D$ is not zero so the recorded counts are not Poisson distributed.

\begin{table*}
\protect\caption{\textbf{PHA with pileup probability distribution parameters.} The
distribution is multivariate normal with the expected value and covariance
shown. In the table, $\lambda$ is the expected number of photons
incident on the detector during the measurement time, $\eta$ is the
expected number of photons arriving during a dead time period, $N_{rec}$
is the total number of photons recorded by the detector in all PHA
bins and $N_{rec,k}$ is the number recorded in PHA bin $k$. \label{tab:PHA-formulas-1}}

\centering{}%
\begin{tabular}{|c|c|c|c|}
\hline 
\multicolumn{1}{|c}{} & no pileup &  & with pileup\tabularnewline
\hline 
\multicolumn{1}{|c}{$\lambda=\int S_{incident}(E)dE$} & $\left\{ 0<E_{1}<\ldots<E_{nbins}\right\} $ &  & \tabularnewline
\hline 
\hline 
photon number spectrum & $S(E)$ &  & $S_{rec}(E)=\lambda p_{rec}(E)$\tabularnewline
\hline 
normalized spectrum & $p(E)=S(E)/\lambda$ &  & $p_{rec}(E)=\sum_{k=0}^{\infty}\frac{\eta^{k}}{k!}e^{-\eta}\left(p^{(k)}*p\right)$\tabularnewline
\hline 
bin probabilities & $P_{k}=\int_{E_{k-1}}^{E_{k}}p(E)dE$ &  & $P_{rec,k}=\int_{E_{k-1}}^{E_{k}}p_{rec}(E)dE$\tabularnewline
\hline 
expected total counts & $\left\langle N\right\rangle =\lambda$ &  & $\left\langle N_{rec}\right\rangle =\frac{\lambda}{1+\eta}$\tabularnewline
\hline 
variance total counts & $var(N)=\lambda$ &  & $var(N_{rec})=\frac{\lambda}{\left(1+\eta\right)^{3}}$\tabularnewline
\hline 
non-Poisson factor & $D=0$ &  & $D=var(N_{rec})-\left\langle N_{rec}\right\rangle $\tabularnewline
\hline 
expected bin counts & $\left\langle N_{k}\right\rangle =\lambda P_{k}$ &  & $\left\langle N_{rec,k}\right\rangle =\left\langle N_{rec}\right\rangle P_{rec,k}$\tabularnewline
\hline 
variance bin counts & $var(N_{k})=\lambda P_{k}$ &  & $var(N_{rec,k})=\left\langle N_{rec}\right\rangle P_{rec,k}+DP_{rec,k}^{2}$\tabularnewline
\hline 
covariance bin counts & $0$ &  & $cov(N_{j},N_{k})_{j\neq k}=P_{rec,j}P_{rec,k}D$\tabularnewline
\hline 
 &  &  & \tabularnewline
\hline 
\end{tabular}
\end{table*}

\noindent{}These formulas were validated using a Monte Carlo simulation\cite{AlvarezSNRwithPileup2014}. 

\textcolor{black}{If the probability of a zero recorded photon count
value is negligible, which is the case with the large expected values
of counts required for material selective imaging, the logarithm of
the data is also normally distributed with parameters \cite{PapoulisAthanasios1965}
\[
\left\langle \mathbf{L}\right\rangle =log(\left\langle \mathbf{N}\right\rangle /\mathbf{N_{0}})
\]
}

\textcolor{black}{
\begin{equation}
var\left(\mathbf{L}\right)=\frac{var\left(\mathbf{N}\right)}{\left\langle \mathbf{N}\right\rangle ^{2}}\label{eq:log-formuals}
\end{equation}
\[
cov\left(\log(N_{1}),\log(N_{2})\right)=\frac{cov(N_{1},N_{2})}{\left\langle N_{1}\right\rangle \left\langle N_{2}\right\rangle }.
\]
}

\subsection{Matrix condition number\label{sub:Matrix-condition-number}}

As discussed in Sec. \ref{sub:CRLB-with-pileup}, $\mathbf{M}$ is
the system matrix at an operating point for the computation of the
$\mathbf{A}$ vector from the measurements, $\mathbf{L}$. The condition
of this matrix measures the size of the perturbation of the solution
$\mathbf{\delta A}$ for a small perturbation of the measurement $\mathbf{\delta L}$.
If the matrix is ill-conditioned then there are large $\mathbf{\delta A}$
perturbations for small $\mathbf{\delta L}$ and the noise in the
A-vector estimates will be large.

The condition of a matrix is discussed in chapter 12 of Trefethen
and Bau\cite{trefethen_numerical_1997}. It is typically quantified
by the condition number, $\kappa$, that is equal to one for a well
conditioned matrix and is large for an ill-conditioned matrix. If
the matrix $\mathbf{M}$ is square and invertible, its condition number
is 
\begin{equation}
\kappa=\left\Vert \mathbf{M}\right\Vert \left\Vert \mathbf{M^{-1}}\right\Vert .\label{eq:cond-sq-M}
\end{equation}

\noindent{}If $\mathbf{M}$ is not square, then the pseudo-inverse
is used for the second factor in Eq. \ref{eq:cond-sq-M}. 

Using the 2-norm, the condition number is equal to the ratio of the
largest and smallest singular values for either a square or rectangular
matrix.

\subsection{Optimal PHA energy bins\label{sub:Optimal-PHA-energy-bins}}

The PHA bins used in the simulation were computed with an algorithm
that maximized the A-vector SNR with the CRLB as the covariance
\[
SNR^{2}=(\mathbf{\delta A)}^{T}\mathbf{C_{A,CRLB}^{-1}\mathbf{(\delta A)}}
\]

\noindent{}The algorithm used as a signal $\mathbf{\delta A}=\left[0,\ 0,\ -1\right]^{T}$.
That is, the imaging task was to detect changes in the third A-vector
components with other components fixed. The results were not changed
for signals resulting from changes to the other components. The SNR
was optimized by exhaustively searching all possible bin widths summing
to the maximum energy in the spectrum with an increment of 3 keV.
The transmitted spectrum for an A-vector in the center of the test
object region described in Sec. \ref{sub:test-object} was used.

\subsection{The test object\label{sub:test-object}}

\begin{figure}
\centering{}\includegraphics[scale=0.6]{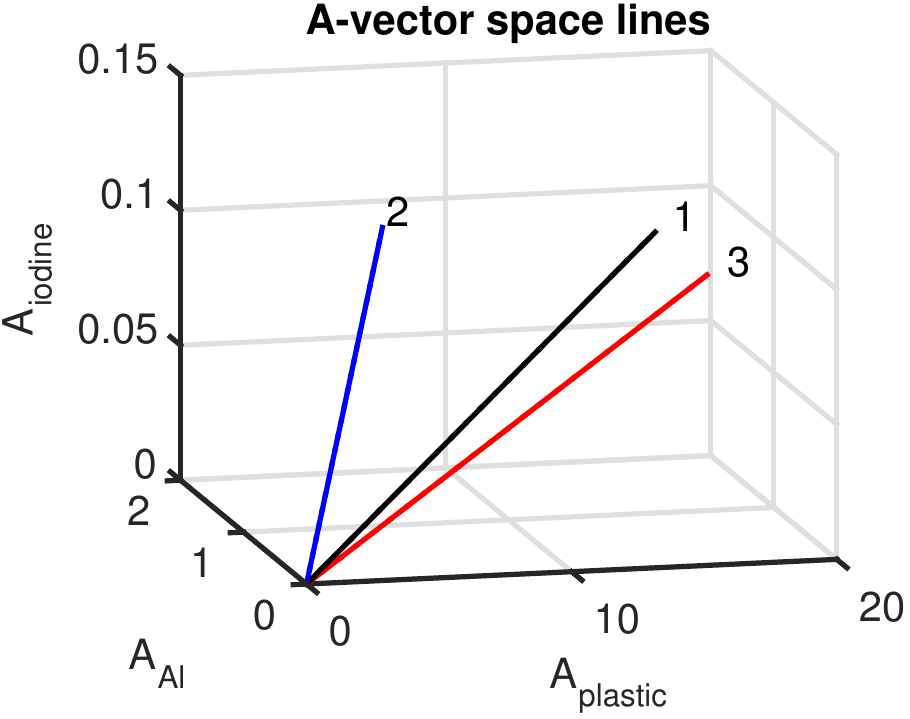}

\protect\caption{Three lines in A-vector space used in the Monte Carlo simulation.
Each line is the A-vectors of different thicknesses of a material
with a specific $\mathbf{a}$ vector of coefficients in its attenuation
coefficient expansion. \label{fig:Three-lines}}
\end{figure}

The performance of the spectral x-ray system was computed for objects
with A-vectors on three lines through the A-vector space as shown
in Fig. \ref{fig:Three-lines}. These correspond to a set of thicknesses
of objects with three compositions. The x-ray attenuation coefficient
of the material of each object is approximated using Eq. \ref{eq:3-func-decomp}.
Summarizing the coefficients as the components of the $\mathbf{a}$
vector, the line integrals are $\mathbf{A=a}W$, where $W$ is the
thickness with units corresponding to attenuation coefficient, for
example $g/cm^{2}$. The A-vectors for each material therefore fall
on a straight line through the origin in A-vector space. The end points
of the lines used in the simulations, which also specify the ratios
of the $\mathbf{a}$ vector coefficients, were $\left[16,\ 1.2,\ 0.1\right]$,
$\left[5,\ 0.9,\ 0.1125\right]$, and $\left[16,\ 0.375,\ 0.1\right]$
$g/cm^{2}$. Fig. \ref{fig:Three-lines} shows a three dimension plot
of the lines.

The test object attenuates the x-ray beam before it is incident on
the detector so the rate of photon arrivals on the detector and therefore
the pileup parameter, $\eta$, decreases as the object thickness increases
along the lines in A-vector space.

\subsection{Compute CRLB for test object\label{sub:Compute-CRLB}}

A 120 kilovolt x-ray tube spectrum was computed using the TASMIP algorithm
of Boone and Seibert\cite{Boone1997}. The number of photons incident
on the object for each detector element or pixel was set to $10^{6}$.
The measurement time was assumed to be 10 milliseconds so the rate
of photons incident on the detector with no object in the beam was
$10^{8}$ photons per second. With detector deadtimes of 10 and 1
nanoseconds the zero object thickness pileup parameter, $\eta_{0}$,
had values of 1 or 0.1. 

The attenuation coefficients of the calibration phantom materials
described in Sec. \ref{sub:test-object} were used as the basis functions
of energy in Eq. \ref{eq:3-func-decomp}. With this choice, the A-vectors
were the thicknesses of each of the materials in the calibration phantom\cite{Alvarez1979}.
The attenuation coefficients of the materials were computed as the
fraction by weight of each element in the chemical formula multiplied
by the attenuation coefficient of that element. The elements' attenuation
coefficients were computed by piece-wise continuous Hermite polynomial
interpolation of the standard Hubbell-Seltzer tables\cite{Hubbell1999}.

For a single A-vector on one of the lines in Fig. \ref{fig:Three-lines},
the TASMIP x-ray tube spectrum and the basis material attenuation
coefficients were used with Eq. \ref{eq:Expected-total-incident}
to compute the spectrum and the expected value of the total number
of transmitted photons incident on the detector sensor during the
measurement time. The sensor was assumed to be perfectly absorbing
so the charge signal of each photon was proportional to the photon
energy. Pulse height analysis was performed for the total energy of
the photons incident on the sensor during the measurement time. As
described in Sec. \ref{sub:Optimal-PHA-energy-bins}, the PHA energy
response functions were computed with an algorithm that maximized
the SNR with no pileup and were assumed to be perfect rectangles.
The CRLB was computed for three and four bin PHA.

The CRLB was computed numerically by approximating the derivatives
in Eq. \ref{eq:F-const-cov} from the first central difference. For
example, to compute $\mathbf{\Delta L}$ we first compute the spectra
through the object with attenuation $\mathbf{A}$ and then with $\mathbf{A+}\Delta\mathbf{A}$.
The transmitted spectra are not affected by pileup since they occur
before the measurement. These transmitted spectra are then used to
compute the expected values of the measurements with pileup using
the formulas in Sec. \ref{sub:PHA-stats}.

\subsection{$\mathbf{M}$ condition number for test object}

As discussed in Sec. \ref{sub:Matrix-condition-number}, we would
expect the estimator noise variance to be large if the $\mathbf{M}$
matrix is ill-conditioned, that is, it has a large condition number.
This was verified by simultaneously plotting the normalized $A_{1}$
variance and the $\mathbf{M}$ condition number along the three lines
in A-vector space.

\subsection{$\mathbf{M}$ condition in 3D A-vector space}

The CRLB and condition numbers showed sharp peaks in plots along the
lines in A-vector space. To better characterize the behavior, the
$\mathbf{M}$ matrix condition number was computed in three dimensional
A-vector space. This was done by computing the $\mathbf{L}$ values
with pileup on a 3D grid of points. The $\mathbf{M}$ matrices at
each of the points were computed using the Matlab $gradient$ function,
which approximates the gradient using the first central difference
of the $\mathbf{L}$ vector data on the neighboring points. The condition
number of the $\mathbf{M}$ at each point was then computed with the
Matlab $cond$ function.

Two dimension data of the condition numbers in the $\left(A_{1},A_{2}\right)$
planes were extracted for each of the $A_{3}$ values. The local maxima
in the 2D condition number data were found using the Matlab $imregionalmax$
function and their three dimension coordinates were saved. A three
dimension plot of the positions of the peak values in the three dimension
A-vector space was then created. 

The pileup parameters $\eta$ on the A-vector space plane of peak
condition numbers were computed.

\subsection{PHA data for large pileup case}

To gain some insight into the cause of the non-invertibility, the
relationship between the peaks in condition number and the count data
in each of the PHA bins were computed as a function of object thickness
by plotting the expected bin count values along the A-vector line
for material 3 (see Fig. \ref{fig:Three-lines}) in separate graphs.

\section{RESULTS}

\subsection{CRLB for three and four bin PHA\label{sub:CRLB-3-4-bin}}

\begin{figure}[h]
\centering{}\includegraphics[scale=0.54]{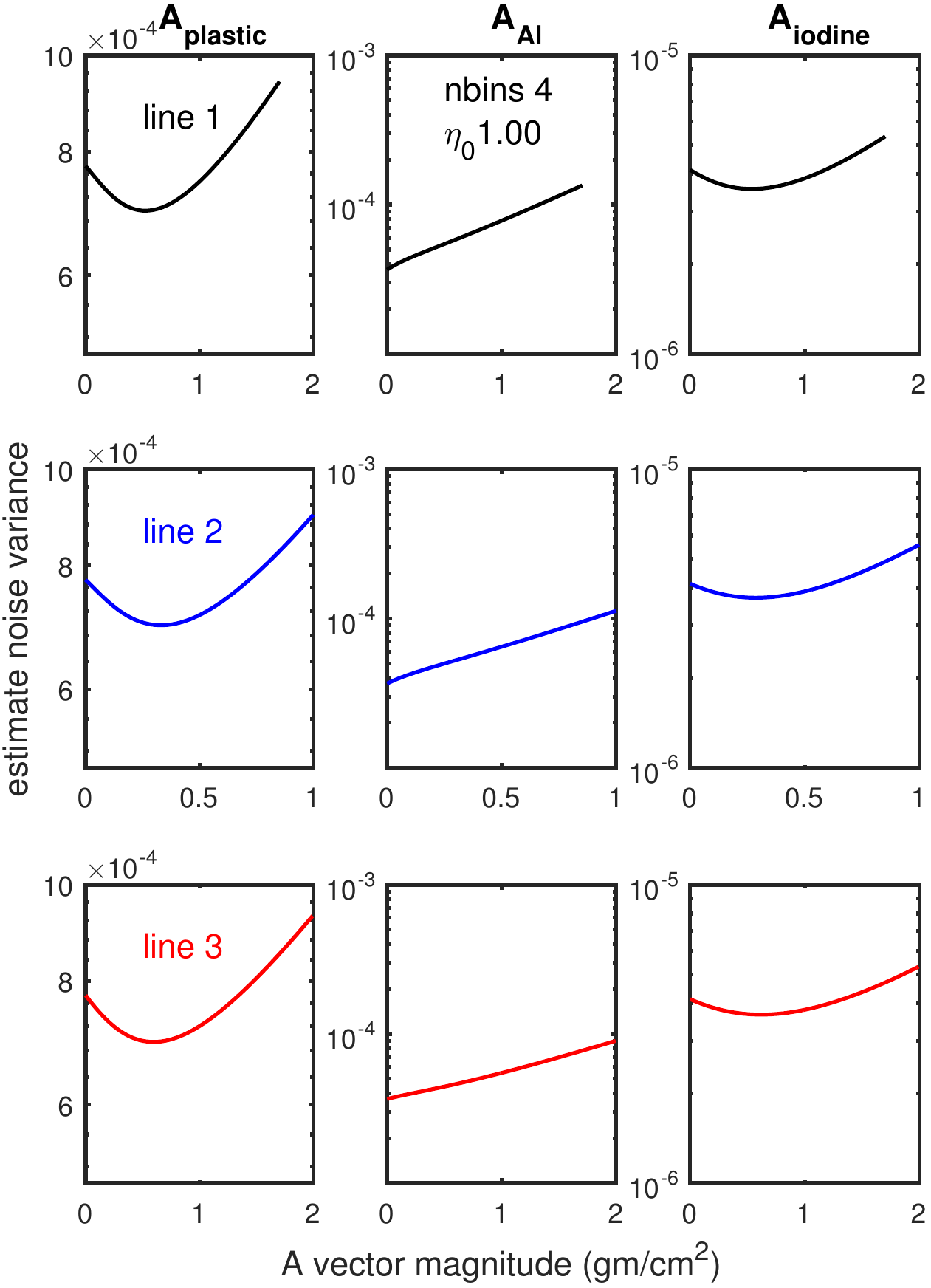}

\protect\caption{Four bin PHA CRLB noise variance as a function of object thickness
along the three lines in A-vector space in Fig. \ref{fig:Three-lines}.
There was a large, zero thickness pileup parameter, $\eta_{0}=1$.\label{fig:4-bin-eta-1}}
\end{figure}

\begin{figure}
\centering{}\includegraphics[scale=0.54]{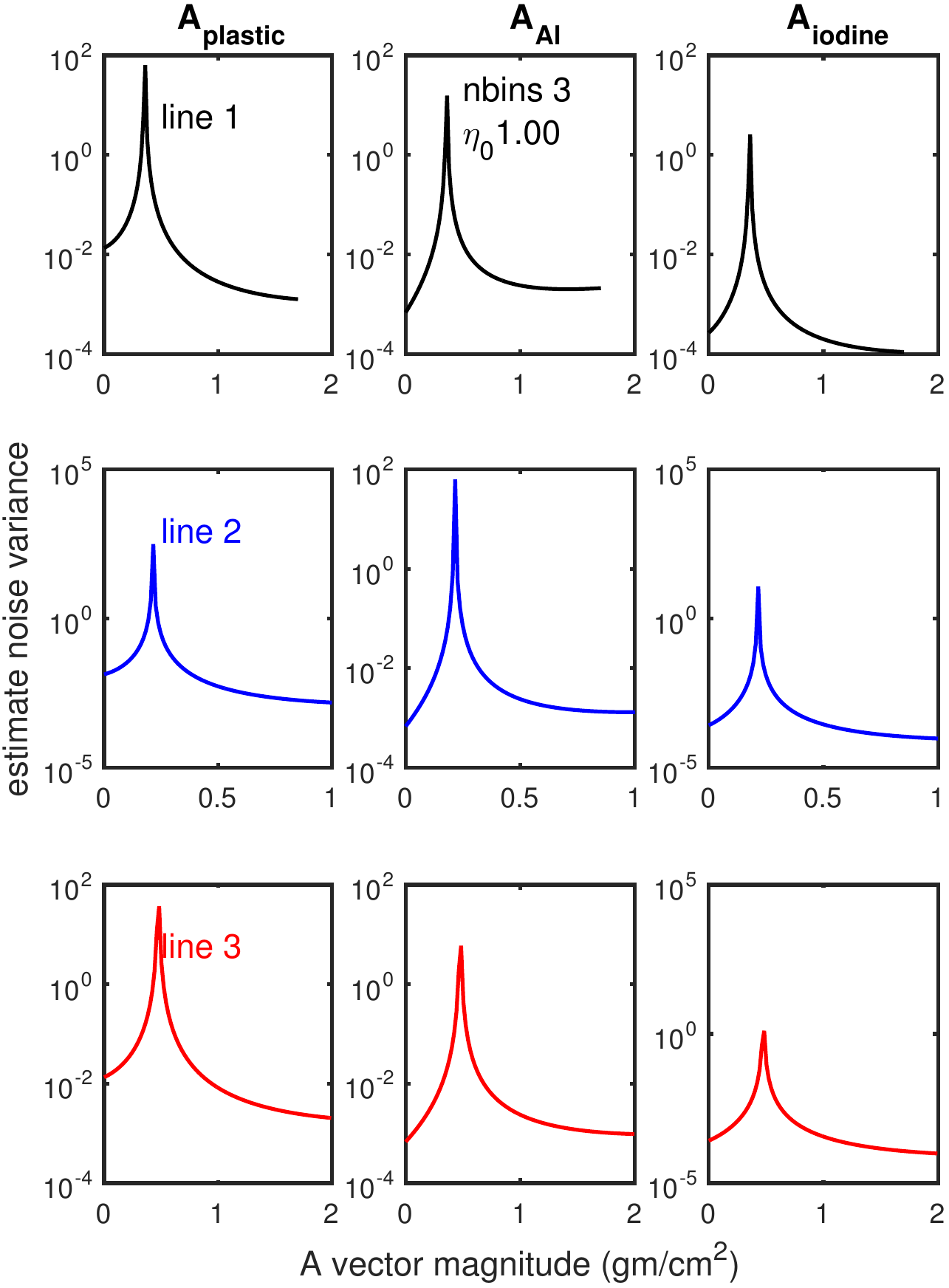}

\protect\caption{Three bin PHA CRLB noise variance with $\eta_{0}=1$ as a function
of object thickness. Note the peaks. They have variance values more
than $10^{5}$ times larger than the 4-bin PHA variance in Fig. \ref{fig:4-bin-eta-1}
for comparable thickness. A logarithmic y-axis scale is used. \label{fig:3-bin-PHA-eta-1}}

\end{figure}

Figures \ref{fig:3-bin-PHA-eta-1} and \ref{fig:3-bin-PHA-eta-1}
show the CRLB noise variance as a function of position along the three
lines in A-vector space in Fig. \ref{fig:Three-lines}. In both figures,
the zero object thickness pileup parameter, $\eta_{0}$, is 1 count
per dead time, a relatively large value. Fig. \ref{fig:4-bin-eta-1}
is for four bin PHA and Fig. \ref{fig:3-bin-PHA-eta-1} is for three
bin.

\begin{figure}[h]
\centering{}\includegraphics[scale=0.54]{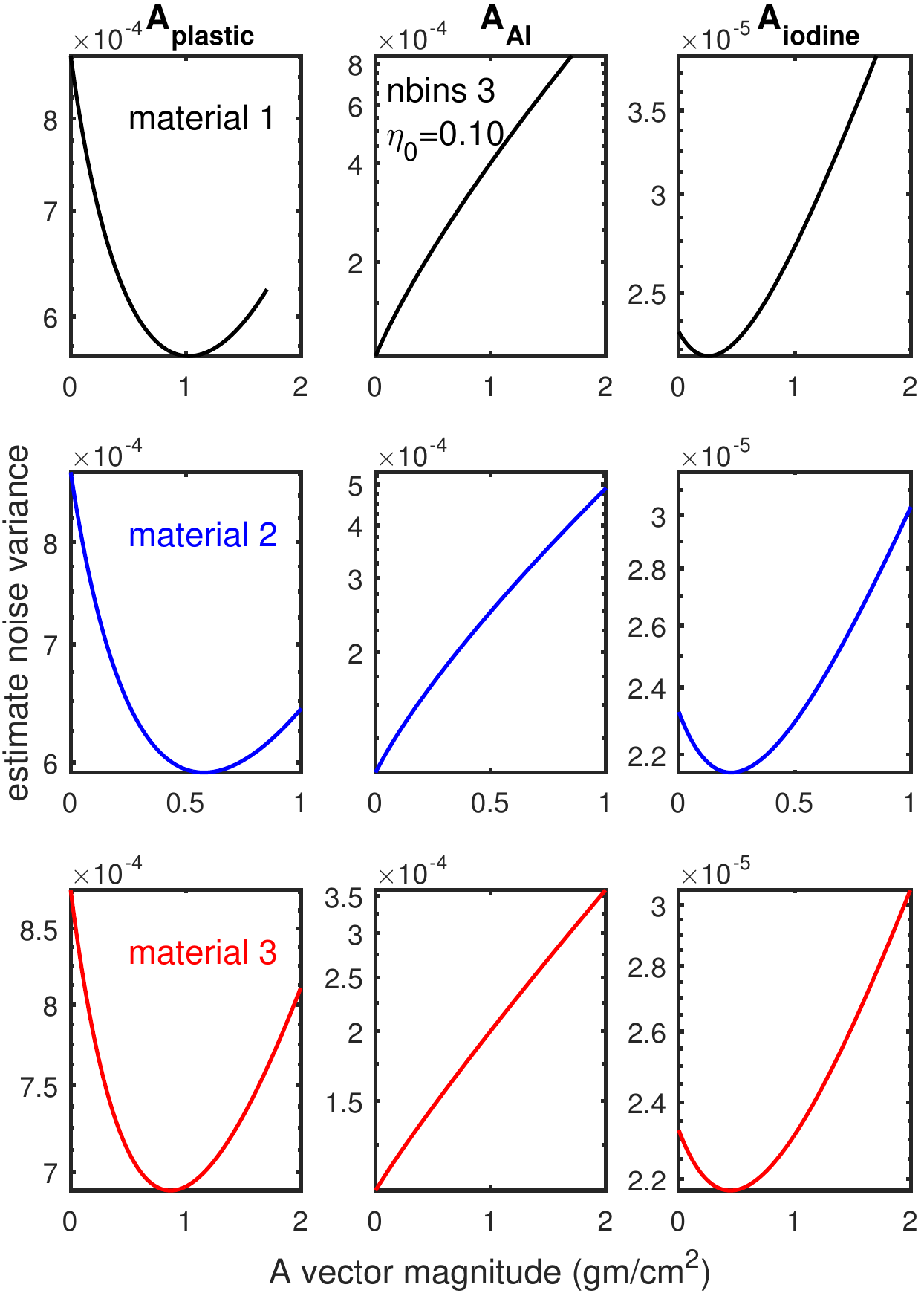}

\protect\caption{Three bin PHA CRLB noise variance with low pileup, $\eta_{0}=0.1$.
Compare with the high pileup case in Fig. \ref{fig:3-bin-PHA-eta-1}.
Notice that with low pileup there are no peaks in the variance. \cite{AlvarezSNRwithPileup2014}.
\label{fig:3-bin-PHA-eta-low} }

\end{figure}

Fig. \ref{fig:3-bin-PHA-eta-low} shows the estimates' CRLB variance
for three bin PHA with a relatively small zero thickness pileup parameter,
$\eta_{0}=0.1$.

\subsection{M condition number and CRLB variance peaks}

\begin{figure}[h]
\centering{}\includegraphics[scale=0.4]{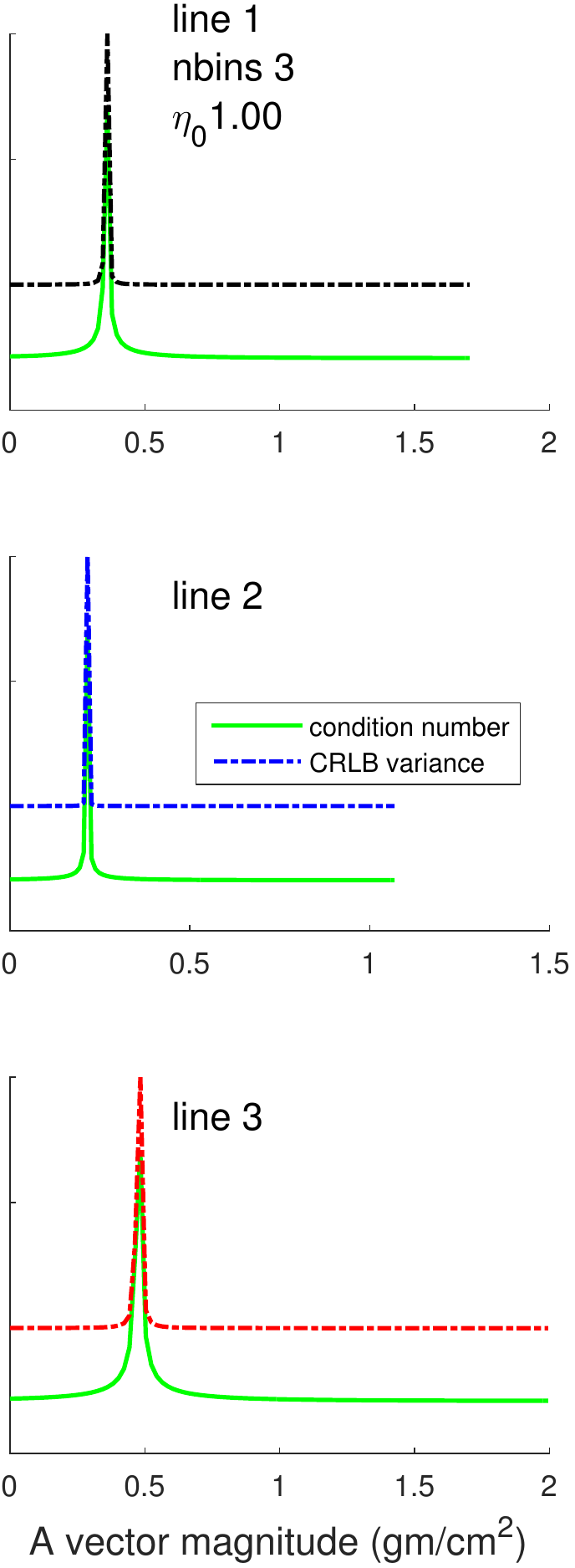}

\protect\caption{CRLB variance peak coincides with M condition number peak. Plotted
is the $A_{1}$ variance and the $\mathbf{M}$ condition number as
a function of object thickness. For visualization, the values are
normalized by dividing by the maximum and offset by subtracting 0.5
from the normalized condition numbers. The y-axis scale is linear.
\label{fig:M-cond-CRLB-peak}}
\end{figure}

Fig. \ref{fig:M-cond-CRLB-peak} shows plots of the $\mathbf{M}$
matrix condition number and the CRLB as a function of position on
line 1. Note that the peaks in variance coincide with a large condition
number i.e. ill-conditioned matrix.

\subsection{Three dimension plot of A-vectors of M condition number peaks\label{sub:3D-plot}}

\begin{figure}[h]
\centering{}\includegraphics[scale=0.45]{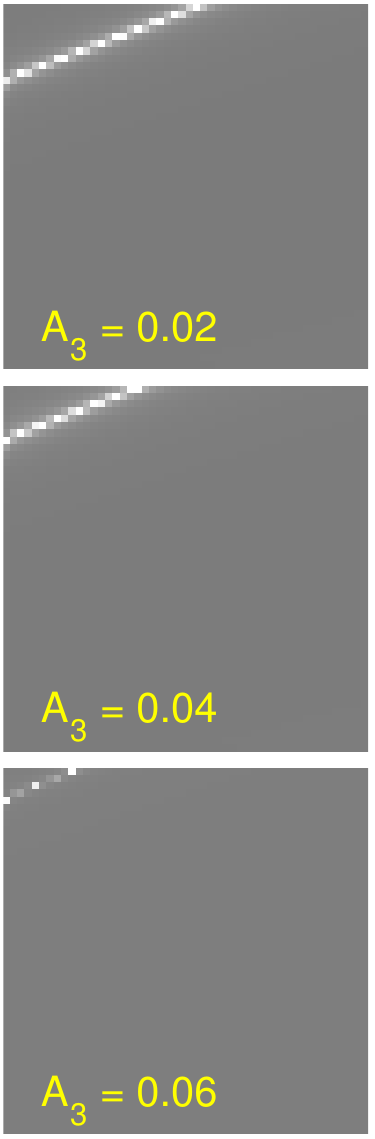}

\protect\caption{Gray scale images of M matrix condition numbers on three planes in
A-space. The images show the $(A_{1},A_{2})$ planes for $A_{3}$
values of .02, .04, and .06 $g/cm^{2}$. \label{fig:Cond-images}}
\end{figure}

Fig. \ref{fig:Cond-images} shows the $\mathbf{M}$ matrix condition
numbers on three planes in A-vector space displayed as gray scale
images. The images show the $(A_{1},A_{2})$ planes for $A_{3}$ values
of .02, .04, and .06 $g/cm^{2}$. 

\begin{figure}[h]
\centering{}\includegraphics[scale=0.54]{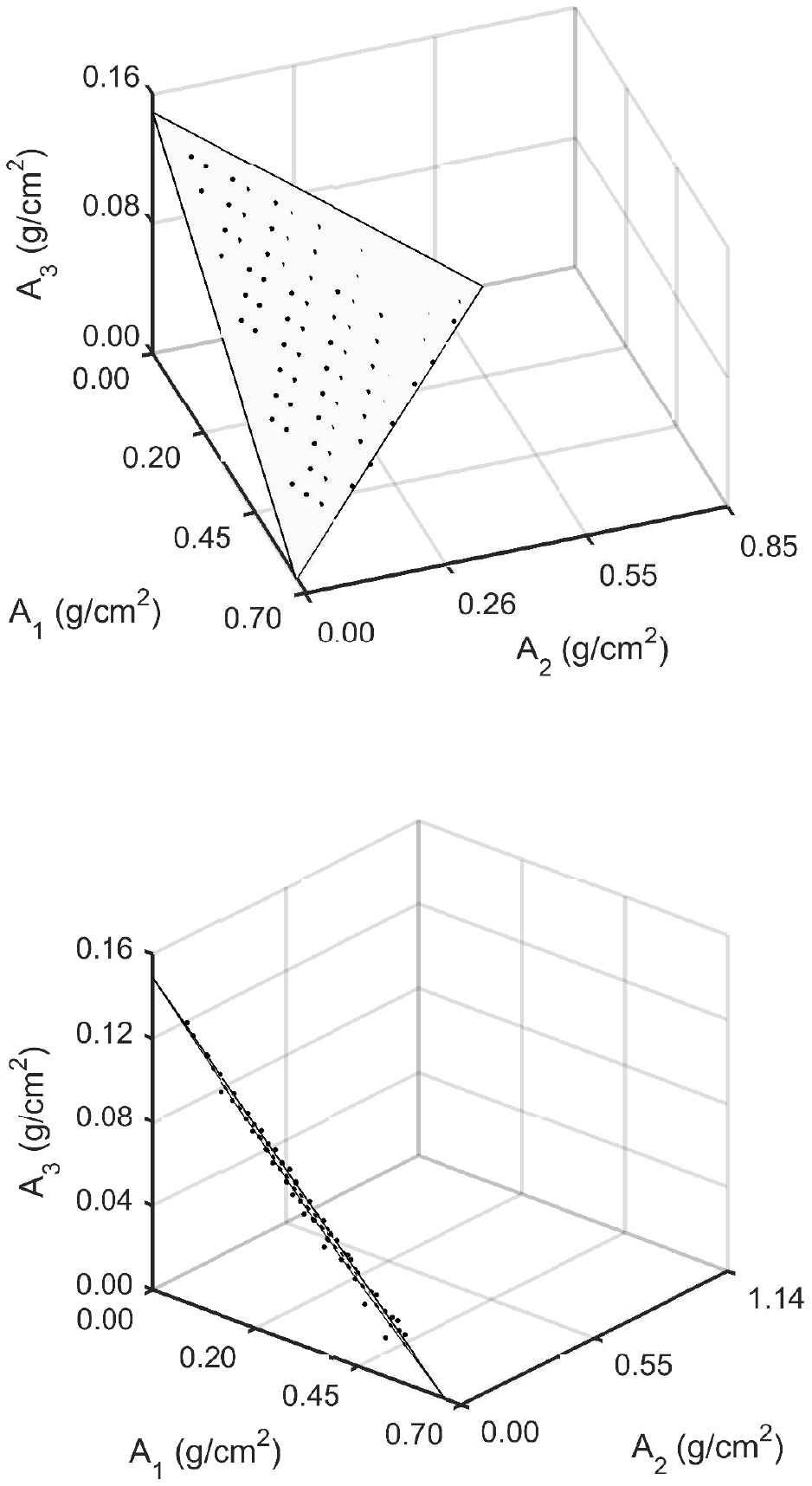}

\protect\caption{Three dimension plot of M condition number peaks. The peaks are the
black dots and the gray triangle is the best fit plane. Two views
of the plane are shown. The top panel shows the peaks and the plane
in a front view. The bottom panel is an edge-on view illustrating
that the peaks are very close to the best fit plane. \label{fig:3D-peaks-plane}}

\end{figure}

Fig. \ref{fig:3D-peaks-plane} shows a 3D plot of the A-vectors of
the peaks detected in the M matrix condition numbers images shown
in Fig. \ref{fig:Cond-images}. There are two views of the best fit
plane. The side view in the bottom panel shows that the peaks are
very close to the plane.

Fig. shows the histogram of the pileup parameter $\eta$ on the plane
of the peak variance in Fig. \ref{fig:3D-peaks-plane}.

\begin{figure}[h]
\centering{}\includegraphics[scale=0.7]{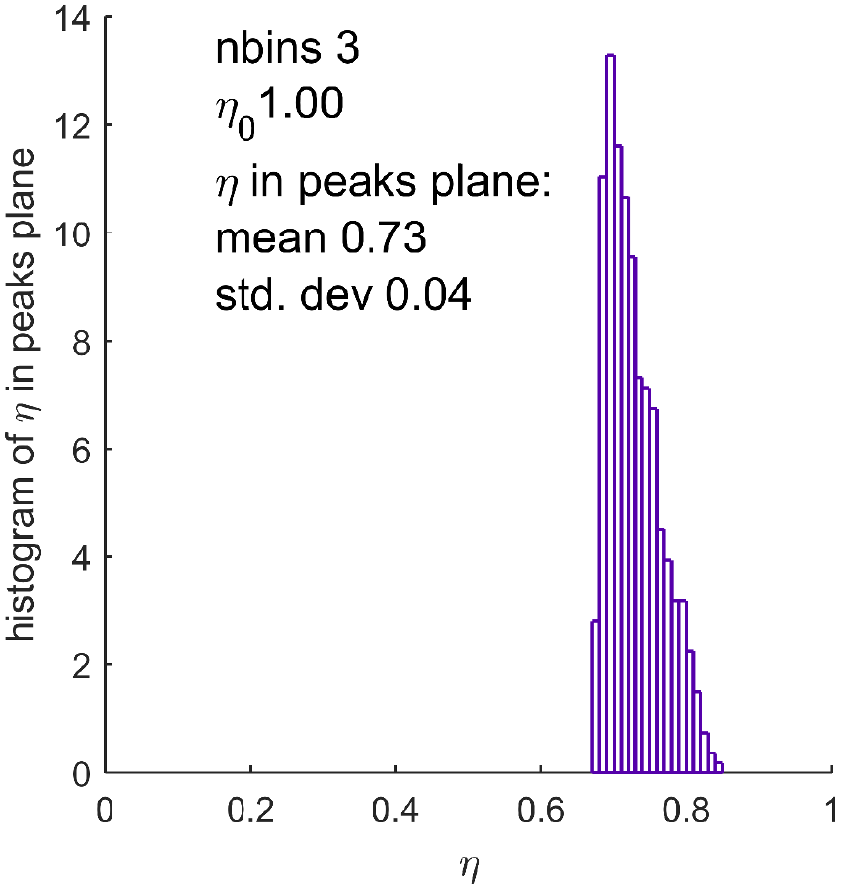}

\protect\caption{Histogram of the pileup parameter $\eta$ on the plane of the peak
variance in Fig. \ref{fig:3D-peaks-plane}.\label{fig:hist-eta-peak-plane}}

\end{figure}

\subsection{PHA bin counts and estimator variance vs. object thickness}

\begin{figure}[h]
\centering{}\includegraphics[scale=0.55]{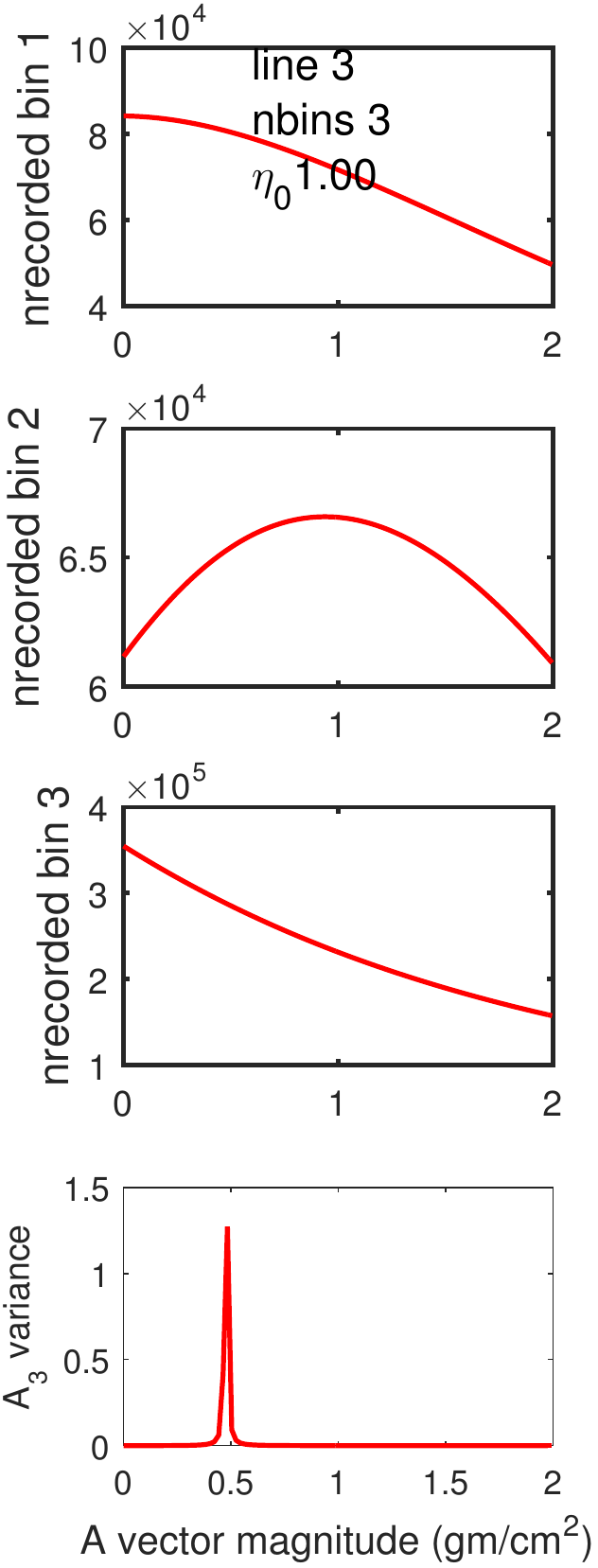}

\protect\caption{The recorded photon counts for each PHA bin and the estimator variance
as a function of the object thickness for line 3 in Fig. \ref{fig:Three-lines}.
The number of PHA bins is 3 and the zero thickness pileup parameter,
$\eta_{0}=1$. The y-axis scale is linear. \label{fig:logIs-with-var}}

\end{figure}

Fig. \ref{fig:logIs-with-var} shows the $\mathbf{L}$ vector components
and the $A_{3}$ estimate variance as a function of the object thickness
for line 3 in Fig. \ref{fig:Three-lines}.

\section{DISCUSSION}

Fig. \ref{fig:3-bin-PHA-eta-1} shows that with large pileup the three
bin PHA data the CRLB variance exhibits a sharp peak that does not
occur with the four bin PHA data results in Fig. \ref{fig:4-bin-eta-1}.
The peak is approximately $10^{5}$ times larger than four bin PHA
variance at comparable object thickness. It is due to large pileup
since the variance with three bin PHA data but with a small pileup
parameter in Fig. \ref{fig:3-bin-PHA-eta-low} does not exhibit the
peak.

Fig. \ref{fig:M-cond-CRLB-peak} shows that the peak in variance coincides
with a peak in the condition number of the $\mathbf{M}$ matrix. The
reason for this is clear from the constant covariance approximation
to the Fisher matrix in Eq. \ref{eq:F-const-cov}, $\mathbf{\left(M^{T}C_{L}^{-1}M\right)^{-1}}$.
If $\mathbf{M}$ is ill-conditioned then the CRLB variance, which
is its inverse, will be large.

Fig. \ref{fig:3-bin-PHA-eta-low} shows that, even though there are
no peaks in variance with low pileup, the variance first decreases
and then goes through a minimum as the object thickness increases.
This is particularly evident in the $A_{plastic}$ component. This
is contrary to the behavior with zero pileup where the variance increases
monotonically with object thickness. With pileup, the pileup parameter
$\eta$ decreases as object thickness increases because the rate of
photon arrivals on the detector decreases. The decrease in pileup
parameter causes the condition number of the $\mathbf{M}$ matrix
to improve so even though the photon counts are decreasing the condition
number improvement compensates and the A-vector variance decreases
for small object thicknesses. As the object thickness increases above
the initial range, the condition number does not decrease rapidly
enough to compensate for the lower photon count and the noise variance
increases.

The figures in Sec. \ref{sub:3D-plot} show that the $\mathbf{M}$
condition number peaks occur on a plane in 3D A-vector space. This
implies that the inverse $\mathbf{L(A)}$ transformation becomes ill-conditioned
for a specific attenuation. The histogram of the pileup parameter
$\eta$ values on the plane in Fig. \ref{fig:hist-eta-peak-plane}
shows that the values are approximately the same, $0.73\pm0.04.$ 

The PHA bin count data for the three bin PHA, high pileup case in
Fig. \ref{fig:logIs-with-var} show that the second bin data are not
invertible since the curve has a zero derivative with respect to the
A-vector magnitude plotted on the x-axis. The shape of the bin counts
curve is due to the fact that, as the object thickness increases,
the count rate and therefore the amount of pileup decreases rapidly.
With less pileup there are fewer cases where the recorded energy is
due to two or more photons. This tends to decrease the relative number
of counts in the third high energy bin and to increase the relative
number of counts in the first and second lower energy bins. 

Notice however that the peak in variance and therefore $\mathbf{M}$
condition number, shown in the bottom panel of Fig. \ref{fig:logIs-with-var}
does not occur at the zero derivative A-vector magnitude value but
somewhat below that value.

High pileup severely distorts the recorded data and reduces its information
content compared with the non-pileup case. The other defects of photon
counting detectors noted in the Introduction also reduce the information
content and it is reasonable that they may also lead to increased
estimator noise variance. Therefore, designers of systems using these
detectors should study the statistical properties of their data and
conduct studies of the implication for the stability of the estimator
output.

\section{CONCLUSION}

Photon counting data with pileup can lead to sharply increased noise
variance in the estimates of the line integrals of the basis set coefficients
in the Alvarez-Macovski method. The increased noise occurs with high
pileup three bin data but does not occur with four bin data or either
three or four bin PHA data with low pileup. The peaks in noise variance
occur for specific object attenuation. State of the art detectors
have pileup as well as other defects so they may also exhibit the
spikes in noise variance found in this study.


\end{document}